\documentclass[aps,twocolumn,superscriptaddress,showpacs]{revtex4}

\usepackage{graphicx}
\usepackage{dcolumn}
\usepackage{amssymb}

\begin{document}

\title{Surface-electronic structure of La(0001) and Lu(0001)}

\author{D.\ Wegner}
\affiliation{Institut f\"ur Experimentalphysik, Freie Universit\"at Berlin, Arnimallee 14, 14195 Berlin-Dahlem, Germany}
\author{A.\ Bauer}
\affiliation{Institut f\"ur Experimentalphysik, Freie Universit\"at Berlin, Arnimallee 14, 14195 Berlin-Dahlem, Germany}
\author{Yu.\ M.\ Koroteev}
\affiliation{Donostia International Physics Center (DIPC), Paseo de Manuel Lardizabal 4, 20018 San Sebasti{\'a}n/Donostia, Basque Country, Spain}
\affiliation{Institute of Strength Physics and Material Science, Russian Acadamy of Sciences, 634021 Tomsk, Russia}
\author{G.\ Bihlmayer}
\affiliation{Institut f\"ur Festk\"orperforschung, Forschungszentrum J\"ulich, D-52425 J\"ulich, Germany}
\author{E.\ V.\ Chulkov}
\affiliation{Donostia International Physics Center (DIPC), Paseo de Manuel Lardizabal 4, 20018 San Sebasti{\'a}n/Donostia, Basque Country, Spain}
\affiliation{Departamento de F{\'i}sica de Materiales and Centro Mixto CSIS-UPV/EHU, Facultad de Ciencias Qu{\'i}micas, UPV/EHU, Apdo. 1072, 20080 San Sebasti{\'a}n/Donostia, Basque Country, Spain}
\author{P.\ M.\ Echenique}
\affiliation{Donostia International Physics Center (DIPC), Paseo de Manuel Lardizabal 4, 20018 San Sebasti{\'a}n/Donostia, Basque Country, Spain}
\affiliation{Departamento de F{\'i}sica de Materiales and Centro Mixto CSIS-UPV/EHU, Facultad de Ciencias Qu{\'i}micas, UPV/EHU, Apdo. 1072, 20080 San Sebasti{\'a}n/Donostia, Basque Country, Spain}
\author{G.\ Kaindl}
\affiliation{Institut f\"ur Experimentalphysik, Freie Universit\"at Berlin, Arnimallee 14, 14195 Berlin-Dahlem, Germany}

\date{November 17, 2005}

\begin{abstract}
Most spectroscopic methods for studying the electronic structure
of metal surfaces have the disadvantage that either only occupied
or only unoccupied states can be probed, and the signal is cut at
the Fermi edge. This leads to significant uncertainties, when
states are very close to the Fermi level. By performing
low-temperature scanning tunneling spectroscopy and \emph{ab
initio} calculations, we study the surface-electronic structure of
La(0001) and Lu(0001), and demonstrate that in this way detailed
information on the surface-electronic structure very close to the
Fermi energy can be derived with high accuracy.
\end{abstract}


\maketitle

%
%

After the first theoretical prediction of a Tamm-like surface
state on Gd(0001)~\cite{Wu91a}, surface states with $d_{z^{2}}$
symmetry around the center of the surface Brillouin zone (BZ) have
been found on all trivalent lanthanide metals by photoemission
(PE) and inverse photoemission
(IPE)~\cite{Li91,Wu91b,Dhe92,Fed94,Bod94,Wes95}. In all cases, the
surface state was found to be located very close to the Fermi
energy, $E_F$, where the accuracy of these experimental methods
deteriorates due to the signal cutoff at $E_F$. This weakness is
manifested in the case of Gd(0001): First, a combined PE and IPE
study concluded that the exchange splitting of the surface state
decreases with increasing temperature and vanishes within
experimental accuracy above the Curie temperature
($T_C$)~\cite{Wes96}; later, a scanning tunneling spectroscopy
(STS) study showed that in fact a constant residual splitting
persists above $T_C$~\cite{Bod98}. This was the first striking
demonstration that STS is a very sensitive technique when states
are located very close to $E_F$, because both occupied and
unoccupied states can be probed within a single measurement,
without any influence on the spectroscopic signal when crossing
$E_F$. Consequently, recent systematic studies of magnetic
lanthanide metals across the whole series of lanthanide metals
demonstrated that STS at low temperatures can probe these surface
states with unprecedented accuracy~\cite{Bau02,Reh03,Bau05}.

Here, we focus on the nonmagnetic 4\emph{f} metals La and Lu. In
both cases, PE and IPE could not clarify whether the binding
energy at the center of the BZ, $\overline{\Gamma}$, is below or
above $E_F$. While for Lu(0001) only PE data
exists~\cite{Wes95,Kai95}, a combined PE and IPE study of La(0001)
concluded that the surface state is partially
occupied~\cite{Fed94}. However, due to the total system resolution
in these experiments, the reliability of this latter statement
must be considered to be questionable. Furthermore, theoretical
band structures are not available for La(0001) and Lu(0001) up to
now. We therefore performed low-temperature STS studies and
band-structure calculations for these two lanthanide surfaces. In
the present work, we compare the theoretical and experimental
results, and we show that -- with the help of the calculated
surface-band structure -- the STS data can be described accurately
by simple models which do not only yield surface-state binding
energies but also lifetime widths and the dispersion of the
surface bands.

%
%

The experiments were performed in an ultrahigh vacuum (UHV)
chamber equipped with a low-temperature STM operated at 10~K
\cite{Bau02}. The samples were prepared \emph{in situ} by vacuum
deposition of electron-beam-evaporated La or Lu metal ($\approx$
30~ML), respectively, on a clean W(110) single crystal substrate
followed by annealing at 800~K in case of La and 1000~K in case of
Lu. STM images were taken in constant-current mode, and the STS
spectra were recorded with fixed tip position,~i.e.\ open feedback
loop. The tunneling current $I$ and the differential conductivity,
$dI/dV$, were recorded as a function of sample bias voltage, $V$,
by modulating $V$ and recording the induced modulation of $I$ via
lock-in technique. A modulation amplitude of 1~mV (rms) at a
frequency of $\approx$ 360~Hz was used, with the time constant of
the lock-in amplifier set between 10 and 100~ms, at a sweep rate
of $\approx$ 2 mV/s. The spectra were taken in both directions,
from lower to higher and from higher to lower sample bias, in
order to verify that energy shifts (due to the finite time
constant) are below 1~meV. Since both the STM tip and the sample
were cooled to 10~K, the energy resolution was $\approx$ 3~meV,
corresponding to $3.5k_{B}T$.

%
%
The calculations were performed using density functional theory
(DFT) in the generalized gradient approximation (GGA) as given by
Perdew and Wang~\cite{Per92}. We use the full-potential linearized
augmented plane-wave method in film geometry~\cite{Wim81,Wei82} as
implemented in the \texttt{FLEUR} code~\cite{FLEUR}. Spin-orbit
coupling is induced self-consistently as described in
Ref.~\onlinecite{Li90}. For a proper description of the 4\emph{f}
electrons, we apply the LDA+U method~\cite{Ani97} in an
implementation similar to that of Shick \emph{et
al.}~\cite{Shi99}. For La, we used values of $U=8.1$~eV and
$J=0.6$~eV, for Lu of $U=4.8$~eV and $J=0.95$~eV. These parameters
were chosen to simulate the experimentally observed positions of
the 4\emph{f} bands~\cite{Lan81}. The La(0001) surface was
simulated by an 11-layer film embedded in semi-infinite vacua, for
the Lu(0001) surface a 12-layer film was used. The vacuum region
was chosen to start 3.2~a.u.\ and 2.9~a.u., respectively, above
the surface atoms of La and Lu. The muffin-tin spheres had a
radius of 3.0~a.u.\ in case of La and 2.8~a.u.\ in case of Lu. A
plane-wave cutoff of $K_{\mathrm{max}}=3.4\, (\mbox{a.u.})^{-1}$
was used, and the irreducible part of the two-dimensional BZ was
sampled with 21 special $k_{\Vert}$ points.
%
%


\begin{figure}
\begin{center}
\includegraphics{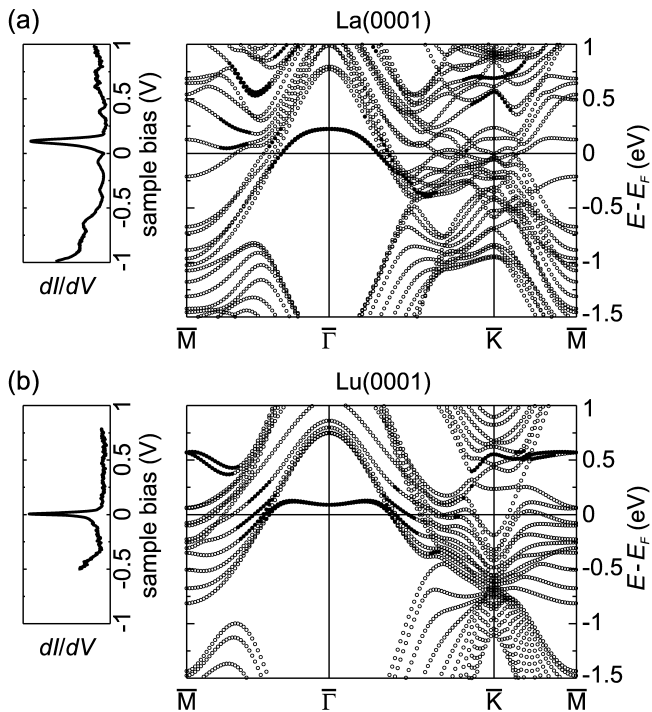}
\caption{\label{fig1} (a)~Tunneling spectrum of La(0001) at
$T=10\;\mbox{K}$ (rotated) and calculated band structure of a
relaxed 11 layer La(0001) film. Surface states are marked by
filled circles, bulk states by open circles. The most pronounced
peak in the STS data at around 0.1~eV is attributed to the narrow
surface-state band around the $\overline{\Gamma}$ point of the
surface Brillouin zone. (b)~Tunneling spectrum of Lu(0001) at
$T=10\;\mbox{K}$ and corresponding band structure of a relaxed 12
layer Lu(0001) film. The surface state appears as a sharp peak at
$E_F$ in STS.}
\end{center}
\end{figure}

In Fig.~\ref{fig1}(a), the STS spectrum of La(0001) is shown
(rotated by 90\textordmasculine), together with the calculated
band structure. The "overview" spectrum is dominated by a narrow
peak at $V \approx 0.1\, \mbox{V}$ which is the signature of the
$d_{z^2}$-like surface state. As STS probes mainly states around
the $\overline{\Gamma}$ point of the projected surface BZ, the
interpretation of the peak as a surface state is clearly confirmed
by the band structure calculation, where a narrow
downward-dispersing surface-state band is clearly visible in the
local gap around $\overline{\Gamma}$ at about 0.2~eV above $E_F$.
Fig.~\ref{fig1}(b) displays the STS data and corresponding
calculated band structure of Lu(0001). Again, the only feature in
STS is a sharp peak located close to $V=0$ (i.e.\ $E=E_F$). It can
be identified as the surface state by comparison with the band
structure, which shows a surface state at 90~meV.

A comparison of the band structures of La(0001) and Lu(0001) with
those of other lanthanide metals shows that their electronic
structures around $\overline{\Gamma}$ are very similar, i.e.\ only
small systematic changes occur across the lanthanide series. A
clear trend in the band dispersion is visible: While for La, the
effective mass is negative (i.e.\ the dispersion is downward
towards lower energies $E-E_F$), Lu first shows a weak upward
dispersion at $\overline{\Gamma}$, which changes to downward with
a band maximum at $\approx 1/4$ way towards the BZ boundary. The
dispersion of Gd is ``in between'' with a very flat surface-state
band~\cite{Kur02}. Calculations for Ce and Tm confirm this
trend~\cite{Sch03,Bod94}. This systematic change further leads to
a decrease of the surface-state bandwidth and higher effective
masses across the lanthanide series. From Fig.~\ref{fig1},
$|m^*/m|>2$ for La and $|m^*/m|>5$ for Lu can be estimated.
Experimentally, in another STS study of Gd, Ho, and Lu we
estimated $|m^*/m|>5$~\cite{Bau02}; a recent PE study of Ce
concludes that $|m^*/m| \approx 7.4$~\cite{Sch03}. These large
effective masses are the result of the high degree of lateral
localization of the Tamm-like lanthanide-surface states.
Consequently, they appear as peaks in STS rather than as step-like
functions (as would be expected for delocalized Shockley-like
surface states, e.g.\ on the (111) surfaces of the noble metals).

For a quantitative analysis of the spectral shape of the surface
states in STS, we have applied a fit analysis that is based on
planar tunneling theory \cite{Wol85}. The fit model was already
introduced in Ref.~\onlinecite{Bau02}. Therefore, we shall only briefly
summarize it here. Assuming that the DOS of the tip is constant
(within the energy range of interest) and that the transmission
coefficient for tunneling, $T(E)$, is bias-independent (for small
$V$), the differential conductivity is given by
\begin{equation}\label{dIdU-alt}
    \frac{dI}{dV} \propto \int\limits_{-\infty}^{+\infty}
    \left(n_{s}T\right)(E)f'(E-eV)dE,
\end{equation}
where $n_s$ is the DOS of the sample, and $f'$ is the derivative
of the Fermi-distribution function; the latter takes into account
the thermal broadening of the spectra due to the finite
temperature in the experiment. For a surface band with quadratic
dispersion,
$E(k_{\parallel})=E_0+(\hbar^2/2m^*)k_{\parallel}^{2}$, and
negative effective mass ($m^*<0$), the DOS is a step function:
$n_s\propto \theta(E_0-E)$. Furthermore, assuming $|m^*| \ll m$
and $E_{\perp}=E-E_{\parallel}<\phi_{\mathrm{eff}}$, $T$ can be
approximated as:
\begin{equation}\label{trans-coeff}
\begin{array}{rcl}
    T(E) & = & \exp\left(-2z\sqrt{\frac{2m}{\hbar^2}(\phi_{\mathrm{eff}}-E_{\perp})}\right)\\
         & = & \exp\left(-2z\sqrt{\frac{2m}{\hbar^2}(\phi_{\mathrm{eff}}-E)+k_{\parallel}^{2}}\right)\\
         & \propto & \exp\left[-p_1(E_0-E)\right],
\end{array}
\end{equation}
where $\phi_{\mathrm{eff}}$ is the effective barrier height, $z$
is the tip-sample distance, and $E_{\perp}$ is the perpendicular
energy component. $p_1$ is a constant parameter, with
$p_1=z\sqrt{2m/\hbar^2\phi_{\mathrm{eff}}}(m^*/m)$ \cite{Bau02}.
Thus, the important term in Eq.~\ref{dIdU-alt}, $n_s T$, is simply
an exponential function that is cut at the band maximum $E_0$ by
the step function.

\begin{figure}
\begin{center}
\includegraphics{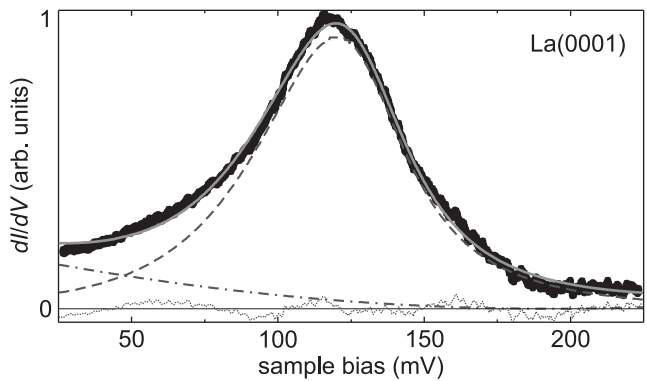}
\caption{\label{fig2}Highly resolved tunneling spectrum of
La(0001) at $T=10\;\mbox{K}$ and least-squares fit (grey line)
composed of a surface-state line according to Eq.~\ref{dIdU-alt}
to \ref{nT-tau} (dashed) and a quadratic background (dash-dotted)
to account for tunneling into bulk states. The residual of the fit
is shown as a thin dotted line.}
\end{center}
\end{figure}

To account for finite lifetimes, each contributing state of the
surface band is broadened by a Lorentzian function. This is
realized by writing $n_s T$ as an integral over Lorentzians
weighted with $T$:
\begin{equation}\label{nT-tau}
\begin{array}{rl}
    n_s T(E) \propto & \int
    \frac{\Gamma(\varepsilon)}{\left(E-\varepsilon\right)^2+\left(\Gamma(\varepsilon)/2\right)^2}\times\\
    & \theta(E_0-\varepsilon)\exp\left[-p_1(E_0-\varepsilon)\right]d\varepsilon.
\end{array}
\end{equation}
$\Gamma$ is the full width at half maximum of the Lorentzian,
which is related to the lifetime $\tau$ by $\Gamma = \hbar/\tau$.
$\Gamma$ is written as a function of energy in order to account
for possible energy dependences of the lifetime due to
electron-electron (\emph{e-e}) and electron-phonon (\emph{e-}ph)
scattering~\cite{Gri81,Qui62,Hel02,Ech04}.

In summary, this model reflects the influence of band dispersion
which leads to slightly asymmetric peaks with broad leading and
narrow trailing edges. Although the above approximations are not
valid for a strongly dispersing band, this model can be
qualitatively conveyed into the model of Li \emph{et al.}, who
analyzed the width of the Ag(111) surface-state
spectrum~\cite{Li98c}. While a large $p_1$ factor (due to a high
effective mass) leads to narrow peaks in STS, a small $m^*$ (as
for the \emph{sp}-like surface states of the noble metals) would
result in a small $p_1$. So the exponential decrease of the STS
signal becomes negligible, and a step function remains which is
broadened by Lorentzians (corresponding to an $\arctan$ function,
as in Ref.~\onlinecite{Li98c}). Indeed, on a larger energy scale also in
tunneling spectra of the surface states on the (111) faces of
noble metals a deviation from a step function can be observed in
the form of a weak decrease of $dI/dV$ towards higher
energies~\cite{Li99,Lim03a}.

\begin{figure}
\begin{center}
\includegraphics{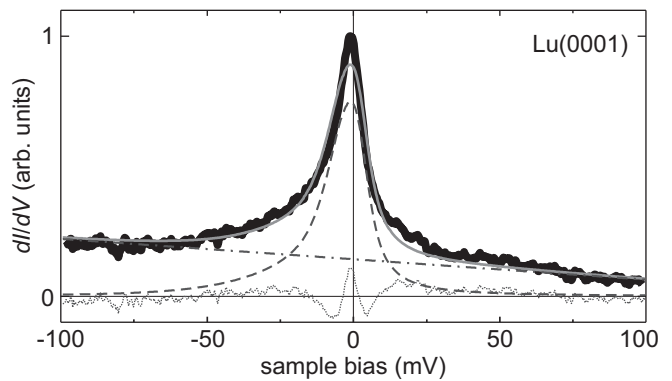}
\caption{\label{fig3}Highly resolved tunneling spectrum of
Lu(0001) at $T=10\;\mbox{K}$ and least-squares fit (grey line)
composed of a surface-state line according to Eq.~\ref{dIdU-alt}
to \ref{nT-tau} (dashed) and a quadratic background (dash-dotted)
to account for tunneling into bulk states. The residual of the fit
is shown as a thin dotted line.}
\end{center}
\end{figure}


In Fig.~\ref{fig2}, a highly resolved tunneling spectrum is
plotted together with the results of the fit analysis. As can be
seen from the residual, the fit describes the experimental curve
very well. This confirms that -- as for Gd, Tb, Dy, Ho, and
Er~\cite{Bau02,Bau05} -- the applied model works convincingly for
a monotonically downward-dispersing surface state with high
effective mass. According to the fit result, the band maximum is
located at $E_0=(130\pm5)\,\mbox{meV}$, and the lifetime width at
the band maximum is $\Gamma(E_0)=(49\pm10)\,\mbox{meV}$, which
corresponds to a lifetime of $\tau(E_0)=(13\pm3)\,\mbox{fs}$.
Thus, the band is clearly unoccupied at $\overline{\Gamma}$ and
crosses the Fermi level at about 1/4 way towards the BZ boundary.
The experimental value for the band maximum is about 70~meV below
the theoretical one. As can be seen in the comparison with
Ce(0001)~\cite{Sch03}, Gd(0001)~\cite{Kur02}, and Lu(0001) (see
below), DFT calculations of the unoccupied lanthanide-surface
states always yield slightly higher energies for the band maximum.

In the following discussion, we compare the results with
literature. A combined PE and IPE study of La(0001) showed that
the surface state is partially occupied at $\overline{\Gamma}$ at
room temperature~\cite{Fed94,Wes98}. As this is clearly not the
case in the present STS study at 10~K, the former observation can
only be explained as a temperature effect in two ways:
(\emph{i})~The surface-state band may shift down with temperature.
However, we have also applied STS on La(0001) at $T \approx 60\,
\mbox{K}$ and did not see any shift of the band maximum.
Therefore, we exclude this possibility. (\emph{ii})~The linewidth
increases with temperature, and thereby the surface state
protrudes over the Fermi edge and becomes partially occupied. Our
analysis of the spectrum at 60~K shows an increase of the
linewidth by $\approx 20\, \mbox{meV}$. In first approximation,
the linewidth increases linearly with temperature due to enhanced
electron-phonon scattering~\cite{Gri81,Hel02}. By extrapolation,
the linewidth at room temperature is estimated to be at least
0.2~eV, i.e. mechanism (\emph{ii}) explains the different results
of this study and of Ref.~\onlinecite{Fed94} as a temperature effect.
Furthermore, taking into account the IPE energy resolution of
$\approx 0.2\, \mbox{eV}$, it is clear that no statement on the
exact position of the band maximum could have been made in
Ref.~\onlinecite{Fed94}. The comparison with our results points out the
advantage of STS, which simultaneously probes both occupied and
unoccupied states around $E_F$.


\begin{figure}
\begin{center}
\includegraphics{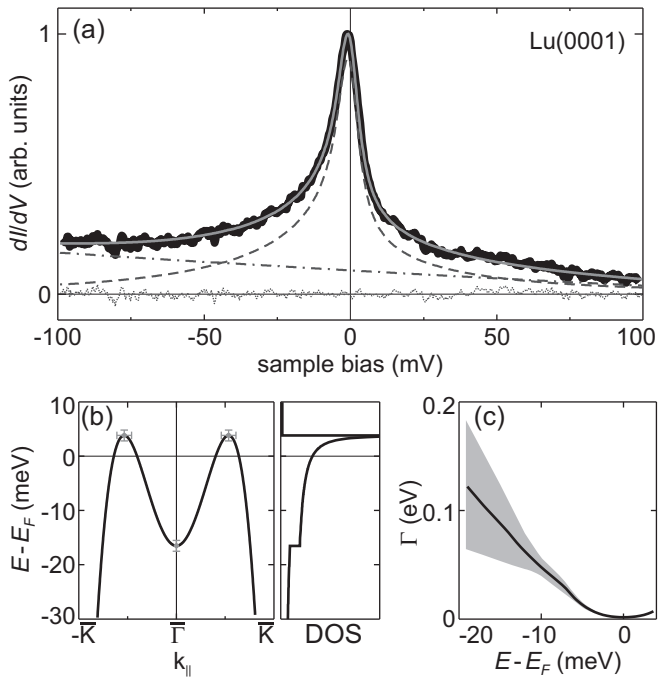}
\caption{\label{fig4}(a)~Same tunneling spectrum of Lu(0001) as in
Fig.~\ref{fig3}. The grey line represents a least-squares fit
composed of a surface-state line according to the new model
(Eq.~\ref{dIdU}) (dashed) and a quadratic background (dash-dotted)
to account for tunneling into bulk states. The residual of the fit
is shown as a thin dotted line. (b)~Dispersion and density of
states (DOS) of the Lu-surface band, and (c)~energy-dependent
lifetime width resulting from the fit analysis.}
\end{center}
\end{figure}

Fig.~\ref{fig3} shows a highly resolved tunneling spectrum of
Lu(0001) plotted together with the least-squares fit results. In
this case, the residual illustrates that the fit model does not
sufficiently well describe the experimental data. As pointed out
recently~\cite{Bau02}, the fit model works only satisfactorily for
Lu(0001), if an additional Gaussian function is added to describe
both the broad tails and the sharp peak in the spectrum. At
present there is, however, no clear physical explanation for such
an additional feature, thus this procedure is questionable. The
model curves in Fig.~\ref{fig3} show the fit result without such a
Gaussian function to emphasize the deviations.

An inspection of the results of the band-structure calculation
[Fig.~\ref{fig1}(b)] explains why the above model, which assumes a
monotonic downward dispersion of the surface band, could not be
successful, because the Lu surface band follows a more complicated
dispersion. The band maximum at $k_{\parallel}\neq 0$ leads to a
singularity in the DOS that must be taken into account, and also
the energy dependence of the tunneling probability is different
for such an M-shaped surface band. Therefore, we introduce a more
sophisticated model to describe the lineshape of the STS data.

The simplest extension of the former model is to assume a
parabolic dispersion of $4^{th}$ order,
$E(k_{\parallel})=E_0+ak_{\parallel}^2+bk_{\parallel}^4$, with
$a>0$ and $b<0$; with this ansatz, $E(k_{\parallel}=0)=E_0$ is the
local band minimum at $\overline{\Gamma}$. The band maximum is
located at $k_{\mathrm{max}}=\sqrt{-a/2b}$ and has the energy
$E_{\mathrm{max}}=E_0+\Delta E$, where $\Delta E = -a^2/4b$. Thus,
$a$ and $b$ can the exchanged by the more descriptive parameters
$k_{\mathrm{max}}$ and $\Delta E$:
\begin{equation}\label{disp-anschaulich}
    E(k_{\parallel})=E_0+2\Delta E (k_{\parallel}/k_{\mathrm{max}})^2-\Delta
    E(k_{\parallel}/k_{\mathrm{max}})^4.
\end{equation}
For $\left| k_{\parallel} \right| \gtrless k_{\mathrm{max}}$, this
results in
\begin{equation}\label{disp-k-E}
    \left| k_{\parallel}(E) \right| =k_{\mathrm{max}} \left(1 \pm \sqrt{1+\frac{E_0-E}{\Delta
    E}}\right)^{1/2},
\end{equation}
and the DOS of the surface band can be calculated as:
\begin{equation}\label{DOS}
    n_s \propto k_{\parallel}\left|\frac{dk_{\parallel}}{dE}\right| =
    \left|\frac{k_{\mathrm{max}}^{2}}{ 4 \Delta E \sqrt{1 + \frac{E_{0}-E}{\Delta E}}}\right|.
\end{equation}

As the dispersion in this model is not simply quadratic, the
approximation for the transmission coefficient in
Eq.~\ref{trans-coeff} is not valid anymore, and the general
formula must be used.

For $E_0<E<E_{\mathrm{max}}$, two solutions exist, i.e.\ for the
same energy there are two contributions to the tunneling current
with different momentum $k_{\parallel}^{(a)}<k_{\mathrm{max}}$ and
$k_{\parallel}^{(b)}>k_{\mathrm{max}}$, respectively. This can be
included by a sum of two transmission coefficients $T_a$ and
$T_b$. For $E<E_0$, $T_a$ is set to zero.

Finally, as for the above model, the broadening of each state due
to the finite lifetime is included by an integral over Lorentzian
functions:
\begin{equation}\label{dIdU}
\begin{array}{rl}
    \frac{dI}{dV}= &
    p_{0}\int\int\frac{\Gamma(\varepsilon)}{\left(E-\varepsilon\right)^{2}+\left(
    \frac{\Gamma(\varepsilon)}{2}\right)^{2}}n_{s}(E)\times\\
    & \left[T_{a}(E)+T_{b}(E)\right]f'(E-eV)d\varepsilon dE.
\end{array}
\end{equation}
In summary, the fit routine possesses the free parameters: $E_0$,
$\Delta E$, $k_{\mathrm{max}}$, $\Gamma$, $\phi_{\mathrm{eff}}$,
and $z$. At a first glance, this number seems to be too large for
a reasonable fit. However, due to the knowledge of the band
dispersion from Fig.~\ref{fig1}(b), $\Delta E$ and
$k_{\mathrm{max}}$ can be restricted to a region that is
compatible with the calculated band structure. Also the last two
parameters can be limited because the barrier height should not
differ much from the work function of Lu ($\approx 3.3\,
\mbox{eV}$), and the typical tip-sample distance is between 5 and
15~\AA.

For $\Gamma$, we take into account the energy dependence due to
\emph{e-}ph scattering ($\Gamma_{e-\mathrm{ph}}$) according to the
Debye model \cite{Gri81,Dou95,Chu03}, and the quadratic energy
dependence due to \emph{e-e} scattering ($\Gamma_{e-e}=2\beta
[(\pi k_B T)^2+(E-E_F)^2]$) as described by Fermi-liquid
theory~\cite{Qui62,Pin69}. $\Gamma_{e-\mathrm{ph}}$ is determined
by the $e-\mathrm{ph}$ mass enhancement parameter $\lambda$ and
the Debye energy $\hbar\omega_D$. $\Gamma_{e-e}$ is described by
the parameter $2\beta$~\cite{Reh03}. Further, an offset $\Gamma_0$
is used as additional parameter in order to include the
possibility that $\Gamma(E_F)>0$. Thus, the lifetime width is only
considered to be energy-dependent in this model but not
$k_{\parallel}$-dependent. We point out that also a momentum
dependence is to be expected, caused e.g.\ by interband \emph{e-e}
scattering with bulk bands, which is more probable when the
excited state is further away from the center of the BZ. However,
including these effects requires a much more complicated
model~\cite{Vit03,Ech04} and is beyond the scope of this study.

Fig.~\ref{fig4}(a) displays the result of a fit analysis of the
same Lu(0001) tunneling spectrum as shown in Fig.~\ref{fig3}, but
now with the new fit model. As can be seen from the residual, the
model curve fits the experimental data perfectly. The results for
the fit parameters are listed in Tab.~\ref{tab1}. In order to test
the reliability of the model, several STS spectra were fitted, and
the reasonability of the parameters was checked. Although these
parameters were treated as free parameters in the fit procedure,
they were found to vary only slightly. Some values that deviate
from the expected values of the calculation will be discussed in
the following. However, the qualitative agreement is good enough
to specify the dispersion of the surface band, cf.\
Fig.~\ref{fig4}(b).

Comparison with the calculated band structure in
Fig.~\ref{fig1}(b) gives good compliance in $\Delta E$, while the
energy position of the band -- as discussed above -- is again
too high in the DFT calculation in GGA approximation. The
difference in $k_{\mathrm{max}}$ is large. While theory puts it to
one fourth, the band maximum in the fit is at about one half of
the surface BZ, where the band structure is already dominated by
bulk bands. Probably, the $k_{\parallel}$ dependence of the fit is
quantitatively incorrect, which is not unexpected since the model
is based on planar tunneling theory. The DOS of the M-shaped
surface band [Fig.~\ref{fig4}(b)], however, shows that -- due to
the singularity at $E_{\mathrm{max}}$ -- states away from
$\overline{\Gamma}$ have a dominant influence on the lineshape of
the tunneling spectrum. Thus, the Lu(0001) surface band is
occupied at the center of the surface BZ, but it has much more
unoccupied DOS slightly above $E_F$. Due to its unique dispersion,
the surface band crosses the Fermi edge twice. The parameters
$\phi_{\mathrm{eff}}$ and $z$ vary barely. A tip-sample distance
of 11 {\AA} is realistic. The effective work function may be
slightly too small but is still in a reasonable regime.

For an estimation of the energy dependence of $\Gamma$, all
relevant parameters have been set free in the fit procedure. The
small offset $\Gamma_0=\Gamma(E_F)=(2\pm1)\,\mbox{meV}$ can be
explained either by a small contribution of defect scattering or
by a poorer experimental resolution (maybe due to electronic
noise). At the band maximum, the lifetime width is
$\Gamma(E_0+\Delta E)=(8 \pm 2)\,\mbox{meV}$. The parameters for
\emph{e}-ph scattering are again acceptable: $\lambda = 1.4\pm0.9$
is higher than the calculated value of Skriver \emph{et
al.}~\cite{Skr90}, but also other experiments find larger
electron-phonon mass enhancement factors up to 2.1~\cite{Wul88}.
The Debye energy, according to the fit result, is $\hbar
\omega_D=(12\pm4)\,\mbox{meV}$, which is between the bulk value of
15.8~meV and the recently determined value for the surface-Debye
energy of 10.4~meV~\cite{Wes95}.

In contrast, the values describing \emph{e-e} scattering are
rather unprecise, particularly for $E_F - E > 10\, \mbox{meV}$.
The $2\beta$ parameter varies between 0 and (unphysical)
100~eV$^{-1}$ for different tunneling spectra. For a clearer
validation, the mean value of $\Gamma$ is plotted versus the
energy [Fig.~\ref{fig4}(c)], the grey-shaded area marks the
uncertainty region. In the relevant region $\pm 10\,\mbox{meV}$
around $E_F$, where the surface-state peak occurs in STS, the
variation of $\Gamma(E)$ is relatively small. Only for lower
energies $E_F - E>10\,\mbox{meV}$, where the contribution of the
surface state to the STS signal decreases rapidly, the
determination of the linewidth becomes untrustworthy, and the
error bars get accordingly very large. In the spectroscopically
crucial region, however, the error bars are smaller than 10~meV.
To further test the model, the $2\beta$ factor was fixed to
realistic values (between 0 and 1~eV$^{-1}$, and the fit procedure
was applied only within the relevant region of the STS spectra.
The resulting parameters do not deviate significantly from those
in Tab.~\ref{tab1}, which emphasizes that the lineshape is
predominantly caused by the unusual DOS of the M-shaped surface
band.


\begin{table}[b]
\caption{\label{tab1}Parameters resulting from the fit analysis of
the Lu(0001) STS data according to the new model. For comparison,
the "starting" values expected from calculations are also listed.}
\begin{ruledtabular}
\begin{tabular}{lcc}
parameters                             & fit result      & "starting" values\\
\hline
$E_{0}$ (meV)                          &  $-17 \pm 2$    & 90\\
$\Delta E$ (meV)                       & $20.3 \pm 0.3$  & 34\\
$k_{\mathrm{max}}$ ($\mbox{\AA}^{-1}$) & $0.64 \pm 0.09$ & 0.27\\
$\phi_{\mathrm{eff}}$ (eV)             & $2.11 \pm 0.05$ & 3.3\\
\emph{z} (\AA)                         & $10.9 \pm 0.5$  & 5...15\\
$\Gamma_0$ (meV)                       & $2 \pm 1$       & 0\\
$\lambda$                              & $1.4 \pm 0.9$   & 0.59\\
$\hbar\omega_D (meV)$                  & $12 \pm 4$      & 15.8\\
$2\beta$ (eV$^{-1}$)                   & 0...100         & $< 1$\\
\end{tabular}
\end{ruledtabular}
\end{table}

We point out again that the difficulties in deriving precise
values for the linewidth may be due to the fact that the model
does not contain a $k_{\parallel}$-dependence of $\Gamma$. For
instance, the \emph{e-e} scattering rate for an excited hole at
the $\overline{\Gamma}$ point, i.e.\ deep within the local band
gap, is predominantly determined by intraband relaxation
processes, while for an excited hole with the same energy $E_0$ at
the decreasing part of the surface band, additional interband
transitions into nearby bulk bands are possible. The interplay
between intra- and interband scattering can be complicated, even
for relatively simple systems as the noble
metals~\cite{Vit03,Ech04}. Also, in case of the far more complex
surface state of Lu(0001), only elaborate calculations will allow
for a quantitative understanding of surface-state dynamics. Given
the current state of the art, the new expanded model can be rated
as a useful description for the STS data of Lu(0001).

%
%

In summary, we have presented low-temperature scanning-tunneling
spectroscopy data and density-functional calculations for the
(0001) surfaces of La and Lu. Fit analyses of the spectra yield
quantitative information about the surface-band dispersions and
the electronic lifetimes of the surface states. The surface state
at the BZ center of La(0001) is clearly unoccupied, but crosses
$E_F$ due to a downward dispersion towards the BZ boundary. The
surface band of Lu exhibits an M-shaped dispersion, with an
occupied local minimum at $\overline{\Gamma}$ and an off-centered
band maximum slightly above $E_F$. We demonstrated that STS
results for Lu(0001) can be described well by an extended -- but
still rather simple -- model. We conclude that STS in combination
with \emph{ab initio} calculations enables one to obtain detailed
information with high accuracy on the surface-electronic structure
of metals near the Fermi energy.

%
%

This work was supported by the  Deutsche Forschungsgemeinschaft
(DFG), projects SfB-290/TPA6 and KA 564/10-1, and by the
University of the Basque Country, Departamento de Educaci\'on del
Gobierno Vasco and MCyT (Grant No.\ MAT 2001-0946). A.B.\
acknowledges support within the Heisenberg program of the DFG.

\bibliography{Wegner_LaLu}

\end{document}